\documentclass[reprint,
superscriptaddress,
amsmath,amssymb,
aps,prd,floatfix,
twocolumn,
showpacs,
]{revtex4-2}

\usepackage{amsthm}
\usepackage{tikz}
\usepackage{graphicx}
\usepackage{dcolumn}
\usepackage{bm}
\usepackage{braket}
\usepackage[english]{babel}
\usepackage{hyperref}
\usepackage{amsmath}
\usepackage{graphicx}
\usepackage{float}
\usepackage{siunitx}
\usepackage{xcolor}
\usepackage{xspace}
\usepackage{eucal}
\usepackage{ulem}
\usepackage{multirow}
\usepackage[utf8]{inputenc}
\usepackage{pgfplots}
\pgfplotsset{compat=1.14}
\usepackage{upgreek}
\usepackage{siunitx}
\usepackage{gensymb}
\usepackage{amsmath}
\usepackage[justification=justified,singlelinecheck=false]{caption}
\usepackage{subcaption}
\usepackage{textcomp}
\usepackage{comment}
\usepackage{cleveref}

\Crefname{figure}{Figure}{Figures}

\usepackage{todonotes}
\DeclareSIUnit\g{g}
\DeclareSIUnit\gal{Gal}
\DeclareSIUnit\torr{Torr}
\DeclareSIUnit\bar{Bar}
\DeclareSIUnit\kelvin{K}
\DeclareSIUnit\inch{inch}
\DeclareSIUnit\joule{J}
\DeclareSIUnit\rad{rad}

\begin{document}

\title{Optomechanical accelerometers for geodesy}

\author{Adam Hines}
\author{Andrea Nelson}
\author{Yanqi Zhang}
\author{Guillermo Valdes}
\author{Jose Sanjuan}
\author{Jeremiah Stoddart}
\author{Felipe Guzman}\email[Electronic mail: ]{felipe@tamu.edu}
\affiliation{Texas A\&M University, Aerospace Engineering \& Physics, College Station, TX 77843}%


\begin{abstract}
We present a novel optomechanical inertial sensor for low-frequency applications and corresponding acceleration measurements. This sensor has a resonant frequency of \SI{4.715\pm0.001}{\Hz}, a mechanical quality factor of \num{4.76\pm0.03e5}, a test mass of \SI{2.6}{\gram}, and a projected noise floor of approximately \SI{5e-11}{\meter\second^{-2}\per\sqrt{\Hz}} at \SI{1}{\Hz}. Such performance, together with its small size, low weight, reduced power consumption, and low susceptibility to environmental variables such as magnetic field or drag conditions makes it an attractive technology for future space geodesy missions. In this paper, we present an experimental demonstration of low-frequency ground seismic noise detection by direct comparison with a commercial seismometer, and data analysis algorithms for the identification, characterization, and correction of several noise sources.
\end{abstract}

\maketitle

\section{Introduction}
Satellite geodesy missions such as the Gravity Recovery and Climate Experiment (GRACE) \cite{https://doi.org/10.1029/2004GL019920} and GRACE Follow-on (GRACE-FO) \cite{kornfeld2019grace} are equipped with low-frequency three-axis accelerometers used to detect nongravitational accelerations. These accelerations include forces such as air drag, thruster activation, and~radiation pressure in the twin-satellite constellation, which are essential measurements to be removed from the missions' gravity observations. The~accelerometers deployed in GRACE and GRACE-FO are electrostatic in nature, reaching acceleration sensitivities of \SI{1e-10}{\meter\second^{-2}\per\sqrt{\Hz}} \cite{flury2008precise,BANDIKOVA2019623}. 

One major challenge posed by these accelerometers, however, is the difficulty to be tested on-ground. These sensors feature quasi-free-falling test masses with no rigid connections to their electrostatic housing. When not in orbit, the~test masses cannot be electrostatically suspended and therefore cannot easily be tested. This problem can hence be avoided by suspending a test mass mechanically, such as with flexures. Optomechanical sensors, for~example, can track the motion of a mechanically suspended test mass using optical readout methods. While a monolithic suspension can help improve significantly the accelerometer robustness, it imposes limitations to its performance as a trend of $1/f^{1/2}$ toward very low~frequencies. 

In spite of such limitations, we expect the sensitivities of these optomechanical devices to be at the levels of valuable scientific observations, while featuring a very small and lightweight footprint. In~addition, their performance is independent of drag conditions (and thus drag-free or drag compensation controls are not required), insensitive to magnetic fields, and~easy to be thermally controlled due to their small size and weight. To~this end, we propose the use of an optomechanical accelerometer for use in future geodesy missions, either by replacing current accelerometers or complementing future baseline instruments as risk-reduction devices, such as gravitational reference sensors~\cite{weber2022application,alvarez2021simplified} or quantum gravimeters and gradiometers~\cite{yu2006development,zhu2022spaceborne,trimeche2019concept}. In~addition, they could be applicable for interplanetary missions studying the gravity fields around other bodies in our solar system~\cite{santoli2020isa}. Finally, they are suitable for ground-based and planetary seismology and geodesy applications~\cite{panning2020deck,fernandez2017volcano}, due to the fact that they can be readily operated under normal gravitational potentials such as Earth's~gravity.

Over the past \SI{10}{years}, optomechanical sensors have been used for many different high-precision sensing applications. Force sensitivities on the order of \SI{10}{\femto\newton} at room temperature have been demonstrated using a \SI{10}{\kilo\Hz} mechanical resonator with a cavity readout~\cite{doi:10.1063/1.4903801}. Optomechanical magnetometers have realized sensitivities better than \SI{1}{\nano\tesla\per\sqrt{\Hz}} over a \SI{}{\giga\Hz} bandwidth using the mechanical modes of barium titanium silicate microspheres~\cite{PhysRevLett.125.147201}. Furthermore, nano-optomechanical displacement sensors with operational bandwidths of several \SI{}{\mega\Hz} have been developed with noise floors on the order of \SI{10}{\femto\meter\per\sqrt{\Hz}} \cite{Lui2020}. Yet other applications include atomic force microscopy and acoustic sensing. Finally, optomechanical sensors have been used for accelerometry, demonstrating an acceleration noise of \SI{100}{\nano\g} over a \SI{10}{\kilo\Hz} bandwidth~\cite{doi:10.1063/1.4881936}.

However, it is worth mentioning that all the above examples use mechanical resonators with eigenfrequencies in the \SI{}{\kilo\Hz} or higher. While low-frequency accelerometers have been developed and tested for various applications, most of them are read out electrostatically~\cite{santoli2020isa}. Hence, optomechanical devices have not been investigated as closely, due to the larger physical dimensions, higher cost, and~poorer sensitivities to high-frequency signals. Yet many fields such as seismology, geophysics, geodesy, hydrology, and~more require the observation of acceleration signals below \SI{1}{\Hz}. Previous work on a low-frequency optomechanical accelerometer demonstrated high mechanical quality factors on the order of \num{1e5} with a \SI{2}{\gram}, \SI{3.7}{\Hz} test mass, corresponding to thermal acceleration sensitivity limits on the order of \SI{1e-10}{\meter\second^{-2}\per\sqrt{\Hz}} \cite{Hines:20}. The~drawback of this device was the inability to integrate the optical readout onto the wafer the resonator was etched from. In this paper, we present our low-frequency optomechanical accelerometer, consisting of a \SI{5}{\Hz} mechanical resonator whose displacement from equilibrium is tracked with laser interferometry. This resonator is designed to be easily integrated with the optical readout, which is advantageous for maintaining a compact form. We describe the characterization process that suggests this resonator is a promising candidate for a novel accelerometer for space geodesy, as~well as ground-based planetary and navigation applications. Using a ringdown measurement, we find the mechanical quality factor of this device to be \num{4.76\pm0.03e5}, corresponding to a thermal acceleration noise on the order of $\SI{5e-11}{\meter \second^{-2}}/\sqrt{f}$ below resonance. Furthermore, we compare acceleration measurements from our resonator to those from a commercial seismometer to demonstrate the resonator's ability to detect seismic motion down to \SI{1}{\milli\Hz}.

\section{Methods}
\subsection{Optomechanical~Accelerometers}
Our optomechanical accelerometer is composed of two main components: the mechanical resonator and the laser interferometric readout. In~this section, we discuss the design of both parts and present characterization measurements which can be used to project the acceleration noise floor of our~device.

\subsubsection*{Resonator---Design and~Characterization}
\label{sec:ResonatorDesign}
\begin{figure}[htbp]
\centering
    \includegraphics[width=0.8\linewidth]{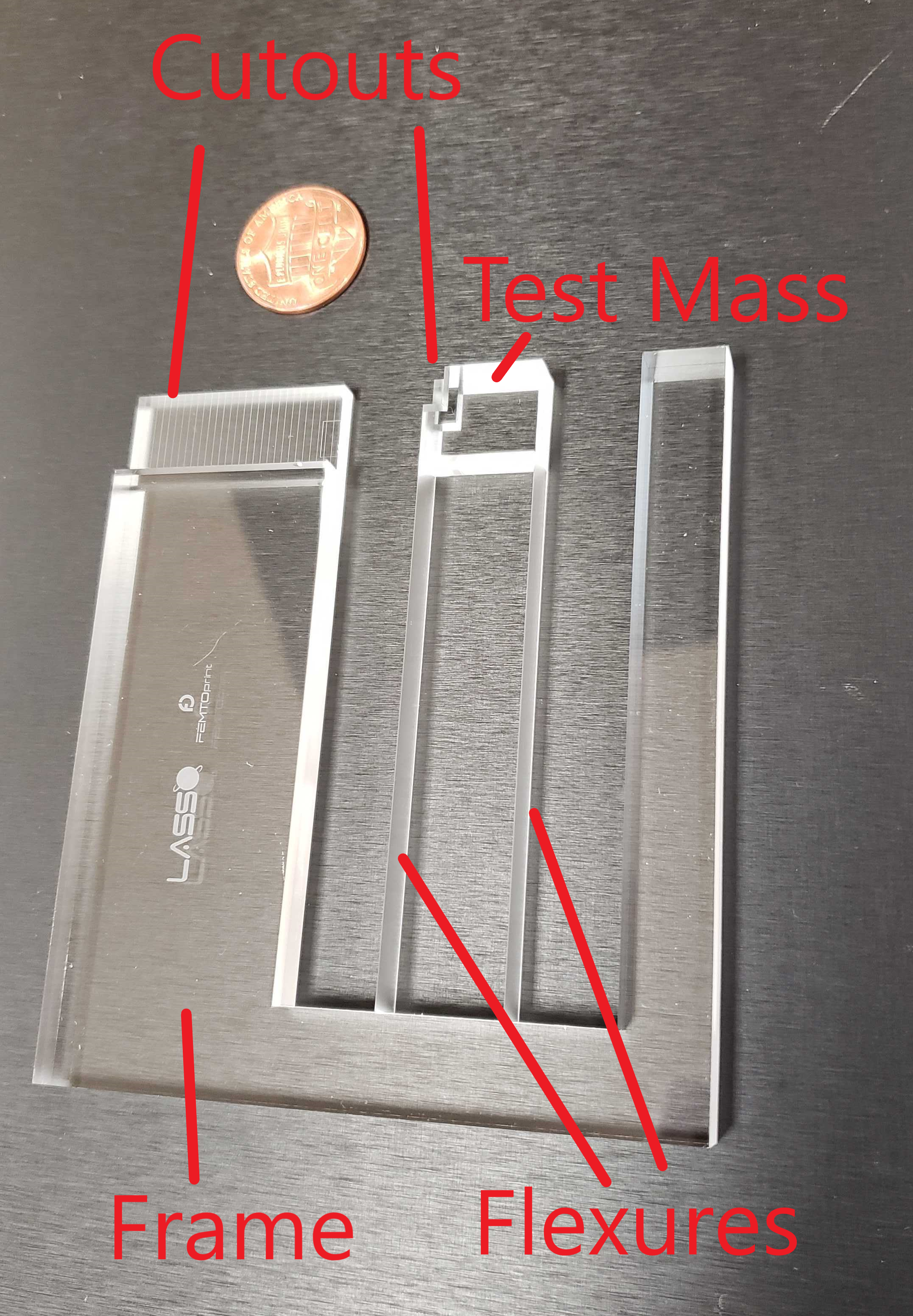}
    \caption{A diagram of our \SI{5}{\Hz} resonator. A~penny is added for~reference.}
    \label{fig:resonator}
\end{figure}

We designed a resonator intended to be used for 1D acceleration measurements, though~a triaxial device would be a straightforward extension of the topology presented here. This resonator was laser-assisted dry-etched from a single monolithic fused silica wafer and was $\SI{90}{\milli\meter}\times\SI{80}{\milli\meter}\times\SI{6.6}{\milli\meter}$ in volume with a total mass of \SI{58.2}{\gram}. Our accelerometer had a smaller form and lighter weight than the GRACE-FO accelerometer, representing a major advantage for satellite missions. An~image of our resonator is shown in Figure~\ref{fig:resonator}. Our design consisted of a \SI{2.6}{\gram} parallelogram test mass supported by two $\SI{100}{\micro\meter}$ thick leaf-spring~flexures. 

A finite element analysis (FEA) simulation performed in COMSOL predicted a \SI{5.5}{\Hz} resonant frequency, which made our resonator suitable for measuring noninertial disturbances below \SI{1}{\Hz}. At~frequencies below resonance, the~test mass response $x$ to an external acceleration $a$  was approximately given by:
\begin{equation}\label{eq:simple_tf}
    x \approx \frac{a}{\omega_{0}^{2}}
\end{equation}
where $\omega_{0}$ is the angular resonant frequency. There is a trade-off between the required displacement sensitivity, how low the resonance can be in a practical resonator, and~its final dimensions. Therefore, a~resonance on the order of \SI{1}{\Hz} allows this low-frequency noise to couple into the test mass motion better than a high resonance device, which in turn relaxes the required test mass readout precision. Conversely, a~lower resonance device requires a larger resonator, which quickly becomes more difficult to work due to the~low stiffness. 

\begin{figure}[htbp]
\centering
\includegraphics[width=0.95\linewidth]{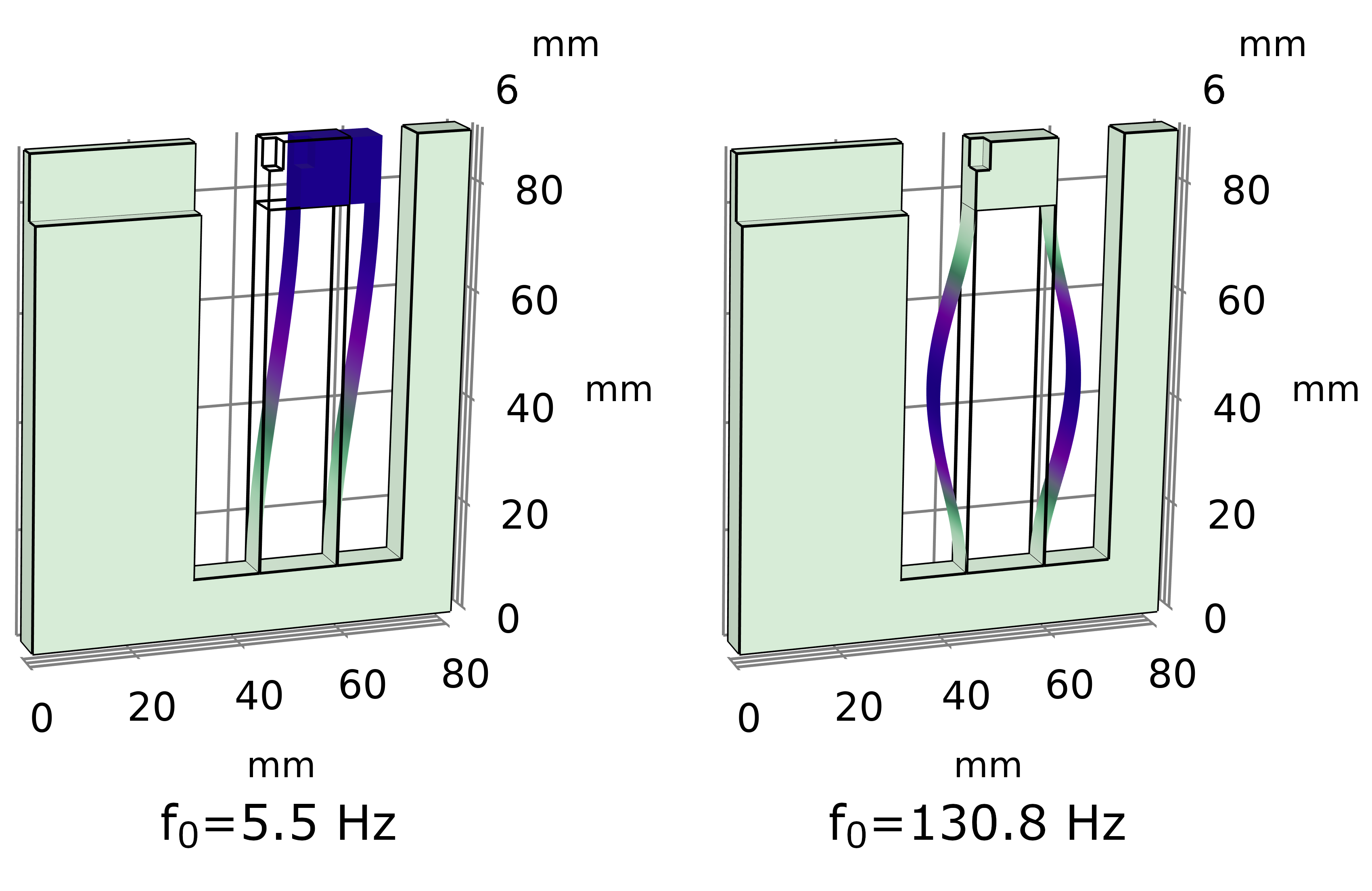}
    \caption{The first two eigenmodes of our \SI{5}{\Hz} resonator modeled in COMSOL. Note that the second mode is larger than the first by a factor of more than 10. Having higher-order modes that are substantially higher than the first mode is desired for minimizing the cross talk between modes observed in~measurements.}
    \label{fig:modes}
\end{figure}

Furthermore, our simulations predicted that all higher-order modes had frequencies above \SI{130}{\Hz}, significantly higher than the lowest resonance by over an order of magnitude. This was desired for mitigating the cross talk between modes that appeared in our measurements, which in turn reduced the noise in our data. Figure~\ref{fig:modes} shows the first two modes of this resonator: the first being the translational mode of the test mass and the second being a violin mode of the flexures with a significantly higher~frequency. 

\begin{figure}[htbp] 
\centering
\includegraphics[width=1.05\linewidth]{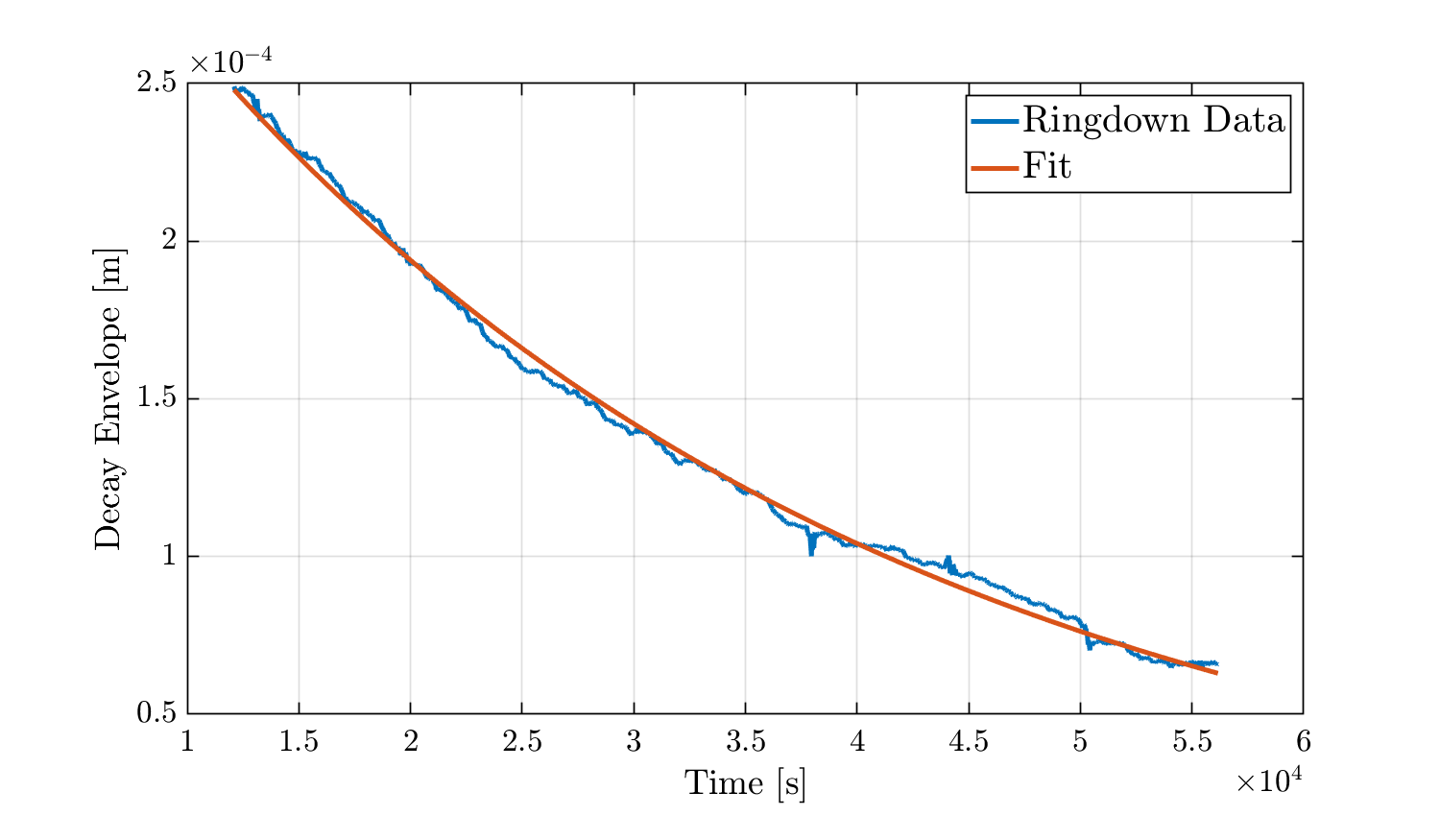}
\caption{Decay envelope of our ringdown measurement fitted to an exponential. The~resonant frequency was removed with a low-pass filter, allowing us to calculate a $Q$-factor of \num{4.76\pm0.03e5}.}
\label{fig:decay-envelope}
\end{figure}

The acceleration experienced by the resonator was recovered by applying a transfer function, which was defined by the resonance and quality factor, to~the test mass's displacement data. This transfer function was given by~\cite{Hines:20}:
\begin{equation}
    \frac{\tilde{x}(\omega)}{\tilde{a}(\omega)} = \frac{-1}{\omega_{0}^{2} - \omega^{2} + i\omega_0\omega/Q} \label{eq:transfer_function}
\end{equation}
where $\tilde{x}(\omega)$ is the test mass motion, $\tilde{a}(\omega)$ is the acceleration coupling into the resonator, $\omega_{0}$ is the resonant frequency, and~$Q$ is the mechanical quality factor. Note that when $\omega << \omega_{0}$, Equation~(\ref{eq:transfer_function}) simplifies to Equation~(\ref{eq:simple_tf}).
To characterize the acceleration sensing capabilities of our resonator, we experimentally determined these parameters using a ringdown test where we deliberately excited the test mass motion and tracked its oscillations. In~the absence of other perturbations, the~oscillation amplitude exponentially decayed, allowing for an easy calculation of $Q$. Figure~\ref{fig:decay-envelope} depicts the decay envelope of our ringdown measurement, which was performed over 14 h at a vacuum pressure of \SI{10}{\micro\torr}. From~this measurement, we found a $Q$-factor of \num{4.76\pm0.03e5} and an $mQ$-product larger than \SI{1250}{\kilo\gram}. Furthermore, taking the fast Fourier transform (FFT) of the raw data yielded a resonant frequency of $f_{0} = \SI{4.715\pm0.001}{\Hz}$, in~good agreement with our simulated value of \SI{5.5}{\Hz}. The~transfer function in Equation~(\ref{eq:transfer_function}) could then be evaluated using our experimentally determined parameters. We applied this transfer function to the displacement measurements in the frequency domain to convert the signal to acceleration~noise.

\subsubsection*{Laser Interferometer---Design and~Characterization}
\label{sec:int}
\begin{figure}[htbp]
\includegraphics[width=\linewidth]{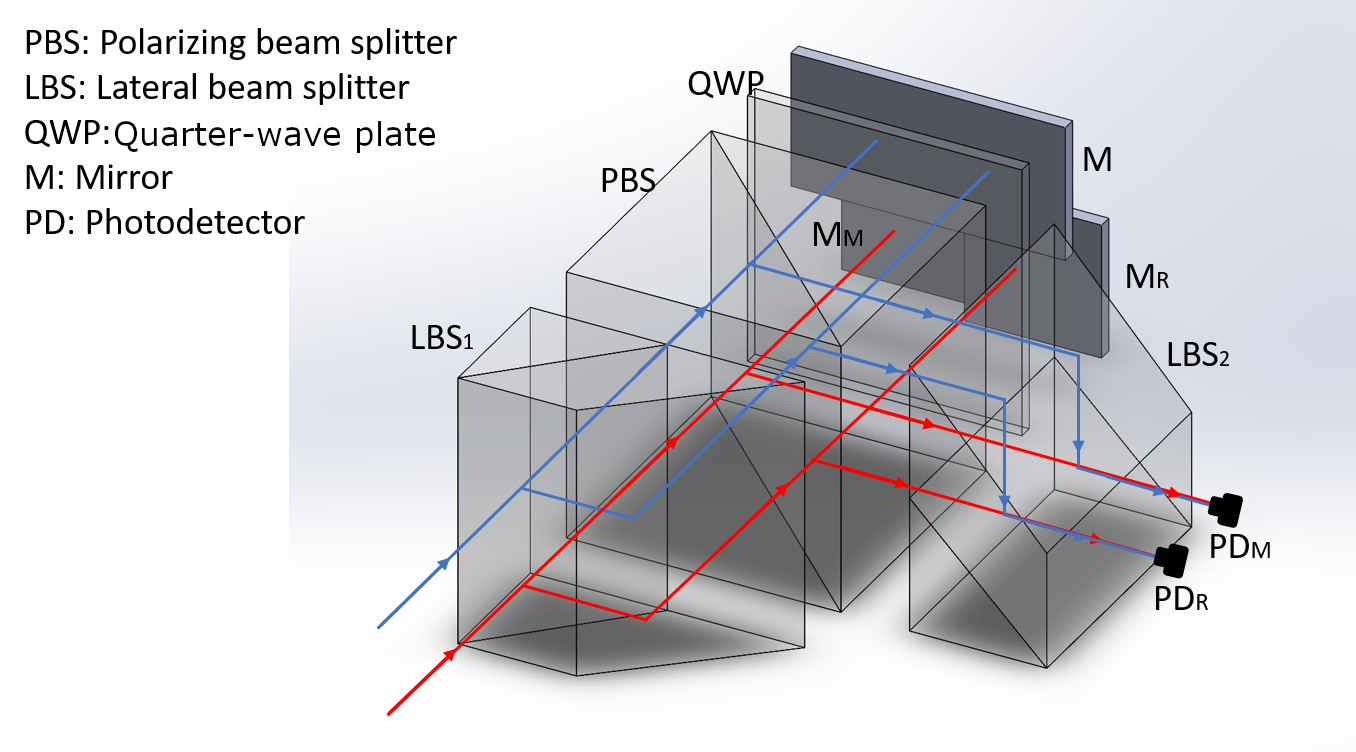}
\caption{A diagram of the heterodyne readout used for our measurements. In~this schematic, the~mirrors $M$ and $M_{R}$ are stationary while $M_{M}$ is the mirror on the test~mass.}
\label{fig:Interferometer}
\end{figure}

In order to measure the local acceleration noise, we must measure the displacement of the test mass from its equilibrium. For~this, we constructed a heterodyne displacement interferometer with a differential phase readout. Cutouts on the test mass and the u-shaped frame of our resonator, shown in Figure~\ref{fig:resonator}, allowed for an easy implementation of mirrors that completed this laser interferometer. The~optical methods and noise rejection schemes used for this readout are discussed in greater detail in~\cite{s21175788}, though~an overview of this optical readout is provided in this section. Figure~\ref{fig:Interferometer} contains a diagram of our laser interferometer. A~fiber-coupled \SI{1064.181}{\nano\meter} beam was split and frequency-shifted by two acoustic-optical modulators (AOMs) to create a \SI{5}{\mega\Hz} heterodyne frequency. The~frequency shifted beams were then injected into a series of prisms that created two distinct Mach--Zehnder type interferometers: one that tracked the test mass motion and one that acted as a~reference. 

The measurement interferometer sent one beam to a mirror on the test mass, $M_{M}$, where the displacement information was imprinted on the phase of the reflected beam. Similarly, the reference interferometer reflected one beam off a stationary mirror, $M_{R}$. The~second beams for both interferometers reflected off the same common mirror, $M$. Common mode noise sources, such as temperature fluctuations in the prisms, coherently affect both interferometers. Therefore, a differential phase measurement rejected some environmental noise and allowed for high-sensitivity displacement~sensing. 

\begin{figure}[htpb] 
\includegraphics[width=\linewidth,trim={0.35cm 0cm 2cm 1.25cm},clip]{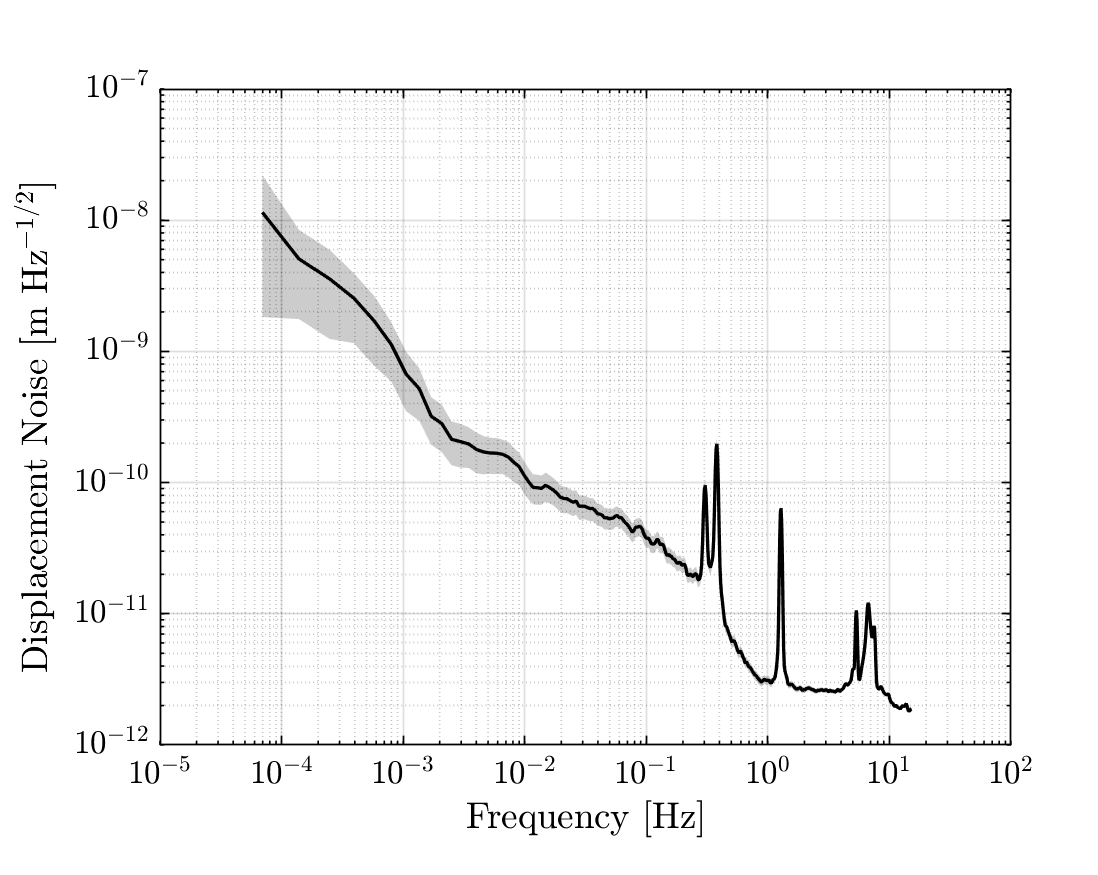}
\caption{A plot of the displacement noise in our heterodyne interferometer. As~this measurement was taken without our resonator, seismic noise was not present in the data. The~shaded area represents the estimated error bars of the~spectrum.}
\label{fig:ifo} 
\end{figure}

Furthermore, the~laser added noise to our data in the form of laser frequency noise. To~combat this, we also introduced a delay-line interferometer to the setup. This interferometer allowed us to make independent measurements of the frequency noise by interfering one beam from the laser with a delayed copy of itself. By~creating a path-length difference of \SI{2}{\meter} in the delay-line interferometer arms, laser frequency fluctuations became the dominant noise source in this fiber interferometer, although~its signal-to-noise ratio (SNR) decreased toward lower frequencies due to fiber noise. With~this measurement, we were able to remove the laser frequency noise from the resonator data in postprocessing. However, for~example, in~future space geodesy missions, a~frequency-stabilized laser source would be available, offering a much better stability than what can be achieved through postcorrection and therefore eliminating the need for a delay-line interferometer. The~readout displacement noise, taken with a stationary mirror in place of the test mass mirror, is shown in Figure~\ref{fig:ifo}. We find that this interferometer can measure displacement data with a sensitivity of \SI{3e-12}{\meter\per\sqrt{\Hz}} at \SI{1}{\Hz} and \SI{7e-10}{\meter\per\sqrt{\Hz}} at \SI{1}{\milli\Hz}. The~peaks near \SI{0.4}{\Hz} and \SI{1}{\Hz} are caused by the mechanical resonances of a vibration isolation platform the interferometer was placed on while~testing.

\subsection{Accelerometer Noise~Floor}
\label{sec:NoiseFloor}
The acceleration sensitivity of our optomechanical accelerometer was anticipated to be limited by both the thermal noise of the resonator and the displacement noise from the readout interferometer. In~this section, we estimated both contributions in order to calculate the acceleration noise~floor.

Thermal noise, which represents the ultimate acceleration sensitivity that can be achieved with our resonator, is caused by gas damping and internal loss mechanisms within the fused silica. These loss mechanisms include bulk losses, surface losses, and~thermoelastic losses. The~thermal acceleration noise can be derived from theory starting from the equation of motion for a resonator, given by~\cite{Hines:20,PhysRevD.42.2437}:
\begin{equation}
    ma = m\ddot{x} + m\Gamma\dot{x} + m\omega_{0}^{2}(1+i\phi(\omega))x \label{eq:eom}
\end{equation}
where $m$ is the test mass, $\Gamma$ is the gas damping rate, and~$\phi(\omega)$ is the internal loss coefficient of our fused silica oscillator. Converting to the frequency-domain and applying the fluctuation--dissipation theorem~\cite{PhysRevD.42.2437}, we find that the thermal acceleration noise is given by:
\begin{equation}
    \tilde{a}_{\mathrm{th}}^{2}(\omega) = \frac{4k_{\mathrm{B}}T}{m\omega}(\omega\Gamma + \omega_{0}^{2}\phi(\omega)) \label{eq:ath}
\end{equation}
where $T$ is the temperature of the test mass and $k_{B}$ is Boltzmann's constant. In~general, both gas damping and internal losses contribute to the thermal noise. However, because~our testing environment could reach \SI{}{\micro\torr} pressures, we operated under the assumption that gas damping was negligible in comparison to internal losses. Previous work on a similar resonator estimated that a vacuum pressure of \SI{10}{\micro\torr} would be sufficient for making gas damping negligible compared to other mechanical losses, which is achieved both in laboratory environments, portable vacuum enclosures, and~certainly space geodesy missions. Therefore, for~the remainder of the paper, we let $\Gamma = 0$. 

We can calculate $\phi(\omega)$ using known equations for the different loss mechanisms~\mbox{\cite{Cumming_2009,Cumming_2012,PENN20063,GRETARSSON2000108}}, but~this is not ideal for our purposes. The~mechanical losses of fused silica have been studied intensively by the gravitational wave community for use in low-thermal noise test mass suspensions. The~models developed by these investigations are useful for optimizing a fused silica sensor's topology, but~the only information we can experimentally obtain for $\phi(\omega)$ is the $Q$-factor. Therefore, we want an equation for the thermal noise in terms of $Q$ rather than $\phi(\omega)$. For~this, we assume that the internal losses are constant in the bandwidth of interest and are given by:
\begin{equation}
    \phi(\omega) = \frac{1}{Q} \label{eq:phi}
\end{equation}

Using these assumptions, we simplify Equation~(\ref{eq:ath}) to:
\begin{equation}
    \tilde{a}_{\mathrm{th}}^{2}(\omega) = \frac{4k_{\mathrm{B}}T\omega_{0}^{2}}{mQ\omega} \label{eq:ath_simp}
\end{equation}

This is the final equation we used to calculate the thermal acceleration noise. Note that larger $Q$-factors lead to lower thermal noise, highlighting the importance of fused silica as a material of choice. Fused silica is known to have very low internal losses at room temperature, with~$Q$-factors well above \num{1e6} for high-frequency resonators~\cite{Numata_2002,doi:10.1063/1.1149159,6782512,7994167,8373358} and above \num{1e5} for low-frequency devices~\cite{Hines:20}. Research into the material properties of fused silica at cryogenic temperatures has shown that the mechanical losses $\phi$ of this material increases substantially as temperature decreases~\cite{schroeter2007mechanical}. At~\SI{30}{\kelvin}, these losses can increase by as much as four orders of magnitude, which would in turn increase the thermal acceleration noise of our optomechanical accelerometer by a factor of 100. Moreover, this work suggested a fused silica resonator could be operated at temperatures as low as \SI{225}{\kelvin} with minimal quality factor degradation of at most a few percent. Hence, we adopted \SI{225}{\kelvin} as the minimum operating temperature of our accelerometer, which was not anticipated to impede our device's performance in the context of space geodesy, as~such systems typically operate near room~temperature.

 In addition to thermal motion, displacement readout noise from the laser interferometer contributed significantly to the acceleration noise floor of our optomechanical accelerometer. For~this section, we note that the Laser Interferometer Space Antenna (LISA) Pathfinder mission has demonstrated an optical readout with a sensitivity reaching \SI{35}{\femto\meter\per\sqrt{Hz}} \cite{PhysRevLett.116.231101}, representing an excellent sensitivity level that can be used to model our own projected noise floor. To~convert this displacement noise to acceleration, we simply applied the transfer function in Equation~(\ref{eq:transfer_function}). This readout noise was assumed to be incoherent with the thermal motion of the resonator, so the two noises added in quadrature:
\begin{equation}
    \tilde{a}_{\mathrm{floor}}^{2}(\omega) = \tilde{a}_{\mathrm{th}}^{2}(\omega) + \big |\frac{\tilde{a}(\omega)}{\tilde{x}(\omega)}\big |^{2}\tilde{x}_{\mathrm{int}}^{2}(\omega) \label{eq:totalnoise}
\end{equation}

\begin{figure*}[htbp]
\centering
\includegraphics[width=\linewidth]{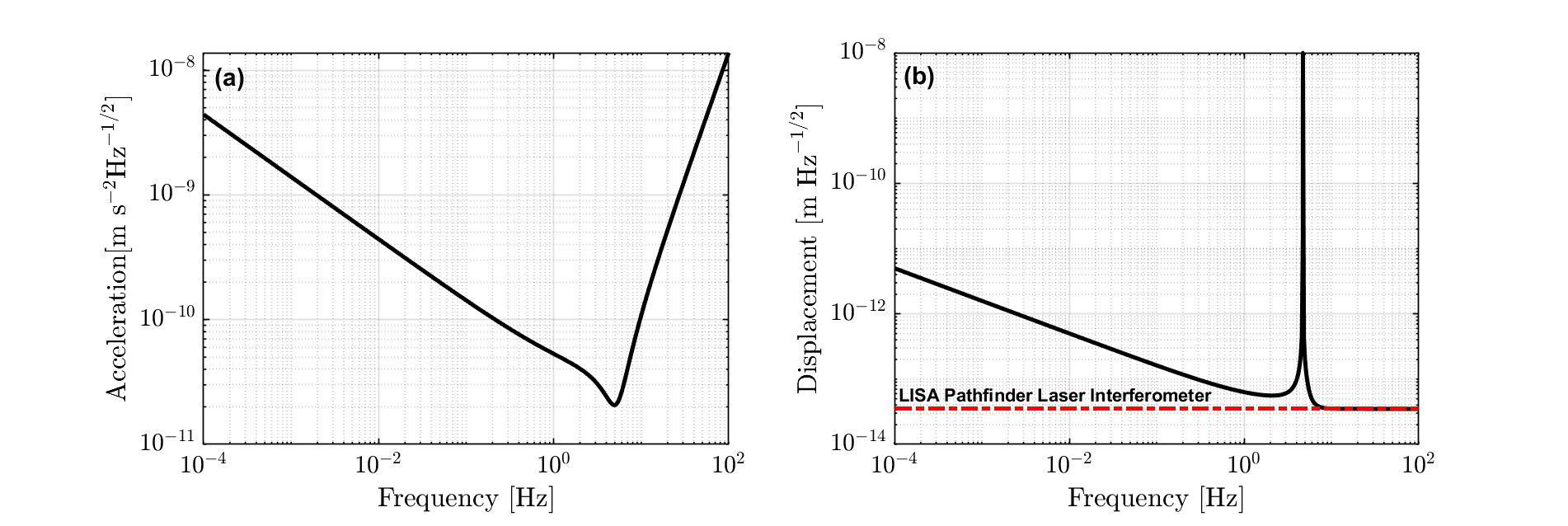}
\caption{(\textbf{a}) The acceleration noise floor of our optomechanical accelerometer including thermal motion from the resonator and readout noise from the laser interferometer. (\textbf{b}) The corresponding displacement readout noise floor required to observe this acceleration noise level. Both plots assume a mechanical $Q$-factor of \num{4.76e5} and a readout noise consistent with that achieved by LISA Pathfinder~\cite{PhysRevLett.116.231101}.}
\label{fig:ThermalNoise}
\end{figure*}

Equation~(\ref{eq:totalnoise}) is plotted in Figure~\ref{fig:ThermalNoise}a, where we observe the acceleration sensitivity is anticipated to be \SI{5e-11}{\meter\second^{-2}\per\sqrt{Hz}} at \SI{1}{\Hz} and increases towards low frequencies as $f^{-1/2}$. Above~resonance, the~noise floor increases rapidly, dominated by readout noise, while below resonance the thermal motion is sufficiently low that the noise floor is \SI{1e-9}{\meter\second^{-2}\per\sqrt{Hz}} at \SI{1}{\milli\Hz}.

\section{Results}
\begin{figure*}[htbp]
\centering
\includegraphics[width=.9\linewidth]{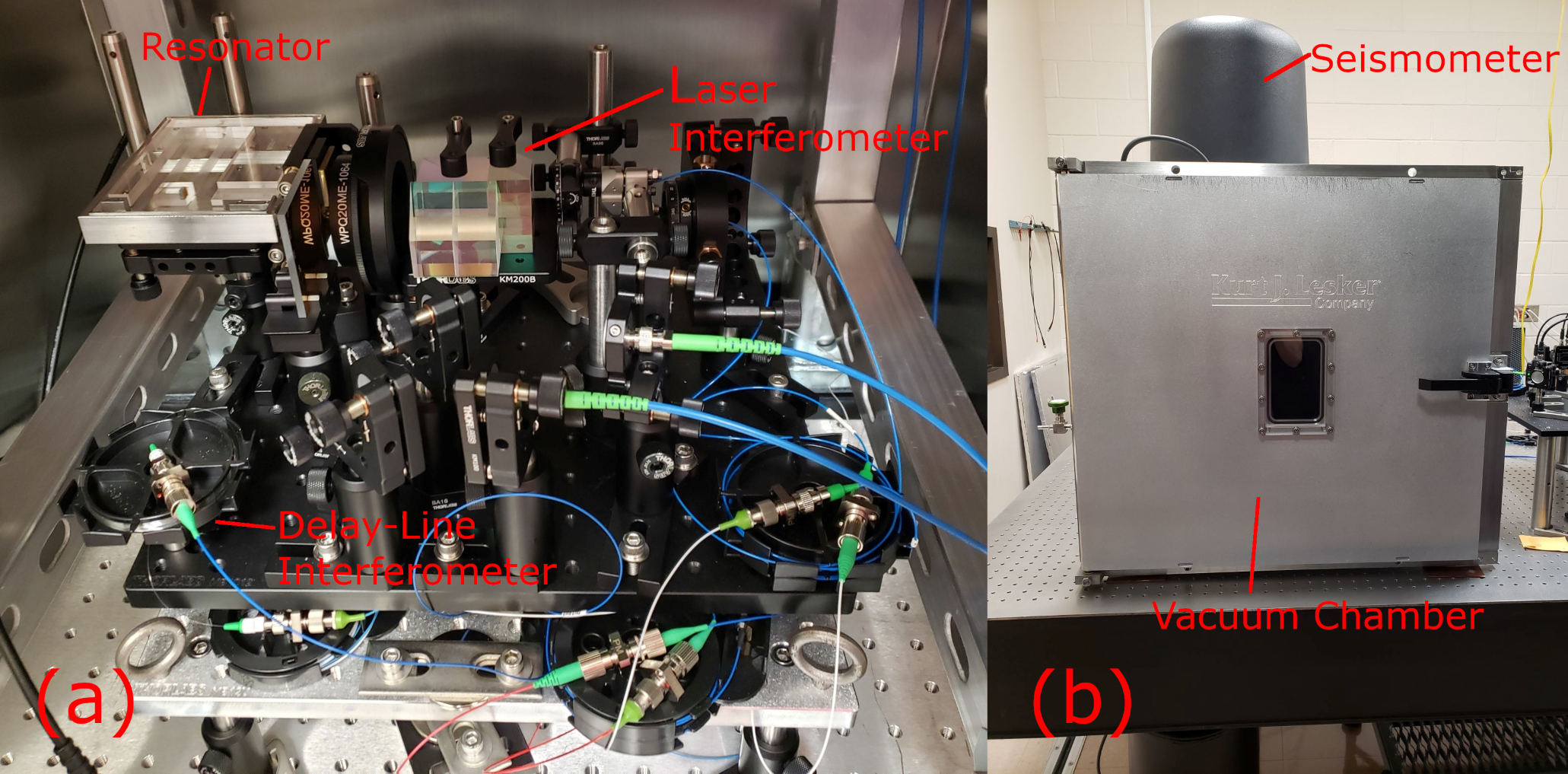}
\caption{(\textbf{a}) An image of our experimental setup, including our fused silica resonator, heterodyne laser interferometer, and~delay-line interferometer. (\textbf{b}) An image of the commercial seismometer's setup relative to our vacuum chamber, which houses the items shown in (\textbf{a}).}
\label{fig:Setup}
\end{figure*}

To demonstrate the acceleration sensing capabilities of our device, we took simultaneous measurements with the optomechanical accelerometer and a Nanometrics Trillium Horizon 120-2 (T120H) seismometer~\cite{TH120}. The~T120H seismometer was placed on top of the vacuum chamber that housed the resonator to ensure seismic noise coupled into both accelerometers coherently. Moreover, the~axis of motion for the resonator was set up to coincide with the x-axis of the seismometer, allowing for an easy comparison of the two sensors. Photos of this setup are shown in Figure~\ref{fig:Setup}. In~this section, we present preliminary measurements taken with our optomechanical accelerometer and describe the correction methods we used to remove various noise~sources.

\subsection{Measuring Seismic~Noise}
\label{sec:Measurement}
\begin{figure}[htbp]
\centering
\includegraphics[width=\linewidth]{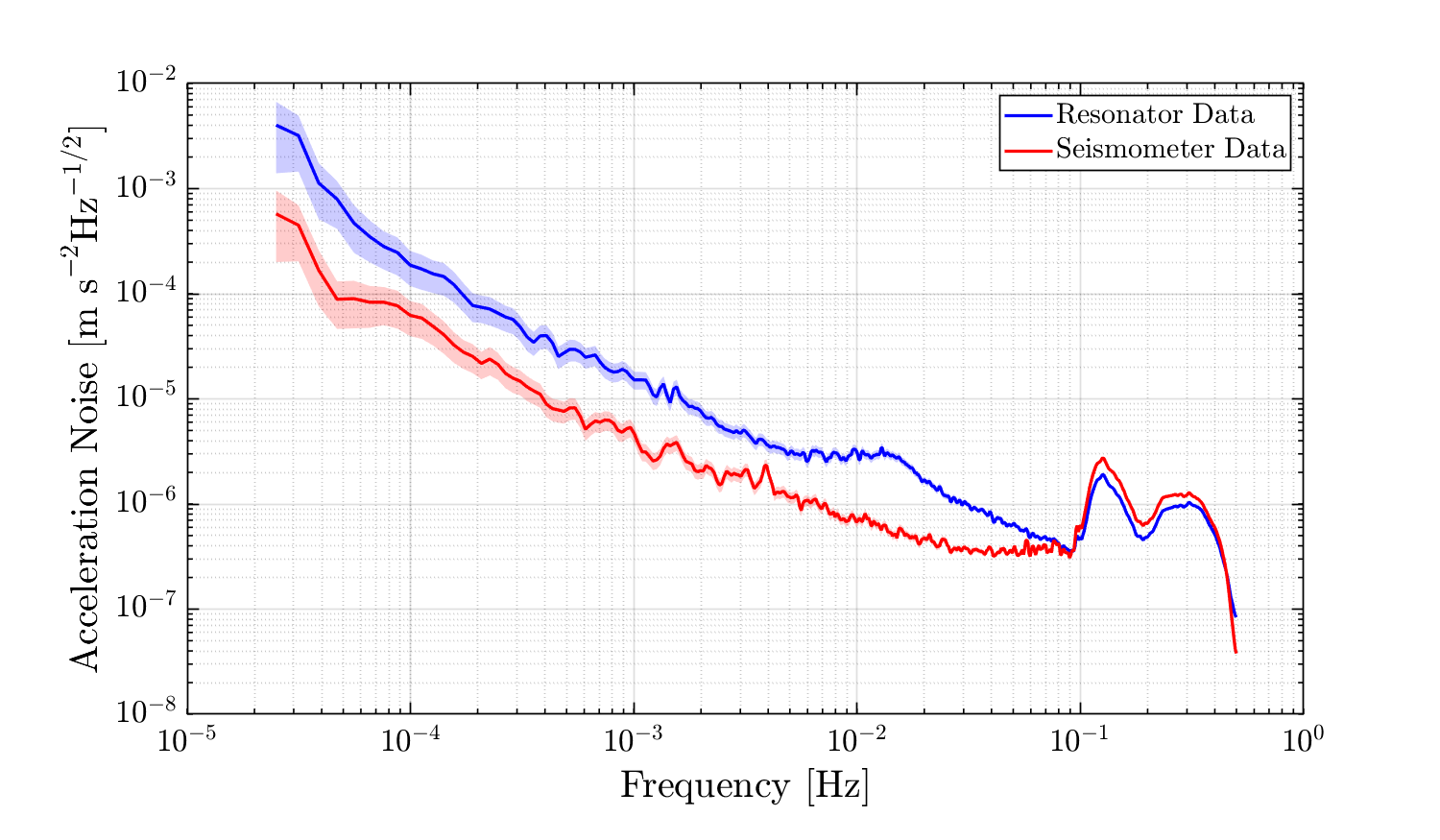}
\caption{The acceleration noise measured by our optomechanical accelerometer and Trillium Horizon 120 seismometer. We observe good agreement between \SI{100}{\milli\Hz} and \SI{500}{\milli\Hz}. The~shaded areas represent the estimated error bars of the~spectra.}
\label{fig:SeismicNoise}
\end{figure}

 Data were taken for 60 h allowing us to observe frequencies as low as \SI{4.6}{\micro\Hz}. The~acceleration noise observed  by both devices is shown in Figure~\ref{fig:SeismicNoise}, where we note the agreement in the microseismic band between \SI{100}{\milli\Hz} and \SI{500}{\milli\Hz}. The~seismometer detected less noise at frequencies below this bandwidth, indicating that our optomechanical accelerometer was limited by environmental noise below \SI{100}{\milli\Hz}. Even though our optomechanical accelerometer's observed noise is significantly higher than the measured optical readout noise in Figure~\ref{fig:ifo}, the~laser path length difference between the measurement and reference interferometers when the resonator is incorporated in the setup is greater than \SI{1}{\cm}. This in turn enlarges the laser frequency noise in comparison to what was observed in the interferometer stability~test.

\subsection{Data~Post-Correction}
The data taken with our optomechanical accelerometer presented in Section~\ref{sec:Measurement} were initially unable to observe the same acceleration noise as our T120H seismometer below \SI{100}{\milli\Hz} due to a combination of signals originating from the environment and the optical readout. However, with~careful environmental monitoring, some of this noise could be removed in postcorrection. In~this section we discuss the different noise removal methods we utilized in our data to achieve a better acceleration sensitivity, and~correspondingly a better agreement with the commercial~seismometer.

\subsubsection*{Time-Domain Linear~Regression}
A portion of the environmental noise in our data could be removed by performing linear fittings to different sets of environmental data taken at the same time as the acceleration measurement. In~addition to the previously discussed delay-line interferometer, this environmental data included the room temperature, vacuum chamber temperature, vacuum pressure, and~the heterodyne signal amplitudes of the measurement, reference, and~delay-line interferometers. Furthermore, we retrieved data on the ambient weather conditions from Texas A\&M University's Research Farm~\cite{TAMUFarm}, as the changing barometric pressure can induce tilting in our optical bench which causes a projection of the local gravity into the resonator's axis of~motion. 

For frequencies below \SI{100}{\micro\Hz}, we found that the barometric pressure dominated the resonator's acceleration noise, with~a coupling coefficient of $\SI{-29.24}{\milli\meter\per\second\squared\per\bar}$ found by linear regression in the time domain. To~isolate the frequency band where the pressure noise dominated, this linear regression was performed on bandpass-filtered data. While effective, the~efficacy of this postcorrection was limited by the data acquisition of the weather station at Texas A\&M's Research Farm, which reported hourly averages. The~ultralow Nyquist frequency of \SI{139}{\micro\Hz} restricted our ability to correct our data for barometric pressure fluctuations above that frequency. As~part of our ongoing efforts to develop this technology, we will include in-house measurements of the ambient laboratory air pressure taken with a commercial barometer sampled at a much higher rate to avoid this~problem. 

The acceleration noise caused by laser frequency noise was found in a similar way. Laser frequency fluctuations appeared in our resonator data as a bump around \SI{12}{\milli\Hz}, which can be observed in Figure~\ref{fig:SeismicNoise}. By~applying a bandpass filter with corner frequencies of \SI{5}{\milli\Hz} and \SI{40}{\milli\Hz} to the resonator and delay-line interferometer data, we found a coupling coefficient of $\SI{1.1421}{\micro\meter\per\second\squared\per\rad}$. This coupling coefficient informed us on how stable a laser would need to be in order for our resonator to be thermally limited. The~phase of our delay-line interferometer, $\phi_{\mathrm{delay}}$, was proportional to our laser source's frequency noise, $\nu_{\mathrm{laser}}$, by~\cite{s21175788}:

\begin{equation}
\label{eq:dlifo}
    \phi_{\mathrm{delay}} = \frac{2\pi\Delta L}{c}\nu_{\mathrm{laser}}
\end{equation}
where $c$ is the speed of light and $\Delta L$ is the path length difference in the delay-line interferometer, which was \SI{2}{\meter} in our case. This suggested that the coupling factor between our resonator's acceleration, $a$, and~the laser frequency noise $\nu_{\mathrm{laser}}$ was \SI{4.709e-14}{\meter\per\second\squared\per\Hz}. Using our projected thermal acceleration noise shown in Figure~\ref{fig:ThermalNoise}, we anticipated a laser frequency stability of $\SI{1.06}{\kilo\Hz}/\sqrt{f}$ would be required for observing the thermal motion of our resonator. This stability is very achievable, and the frequency stability of the laser onboard GRACE-FO is much better than this requirement~\cite{thompson2011flight}.

\subsubsection*{Transfer~Function}
The linear regression method of noise subtraction that we used for the barometric pressure and laser frequency corrections assumed that the phase and amplitude of their respective coupling coefficients were uniform over the bandwidths we were correcting. However, noise generally does not couple into a given system uniformly in frequency; both the amplitude and phase of the coupling factor can be frequency-dependent. The~temperature inside our vacuum chamber for example was a significant source of noise in our experiment, but~it did not couple uniformly in frequency. Because~of this, we had to use a different approach to remove the temperature effects from our data. We corrected for the vacuum chamber temperature by calculating a transfer function, $H_{Ta}(\omega)$, between~the temperature and acceleration data~\cite{nofrarias2013subtraction}. This transfer function is typically estimated by taking the ratio of the cross-power spectral density (CPSD) of the acceleration and temperature data, $S_{Ta}(\omega)$, and~the power spectral density (PSD) of the acceleration, $S_{aa}(\omega)$ \cite{bendat1978statistical}:
\begin{equation}
    H_{Ta}(\omega) = \frac{S_{Ta}(\omega)}{S_{aa}(\omega)} \label{eq:H}
\end{equation} 

For this spectral analysis, we used a Nuttall window function and \num{1.98e5} samples. The~number of averages increased with frequency. The~lowest frequency bin, \SI{5}{\micro\Hz}, had only one average, which increased to over 1100 averages at \SI{0.5}{\Hz}. At~\SI{230}{\micro\Hz}, there were 10 averages, and~so we adopted this frequency as the approximate cutoff below which the spectral analysis had too few averages to be reliable. This analysis was performed with LTPDA, an~open-source MATLAB toolbox for data analysis and signal processing developed and distributed by the LISA Pathfinder community~\cite{LTPDA}.

\begin{figure}[htbp] 
\centering
\includegraphics[width=\linewidth]{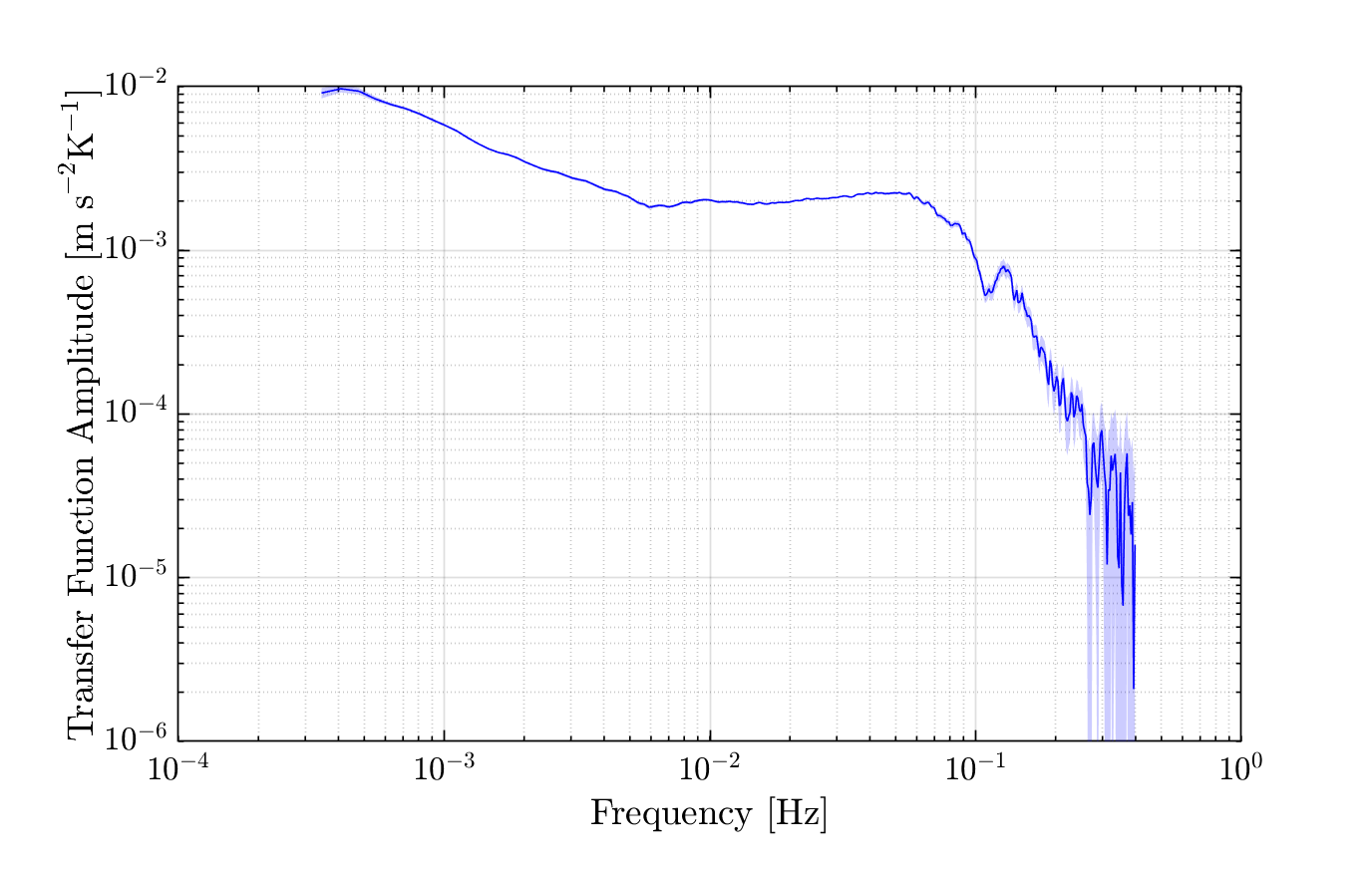}
\caption{The transfer function between vacuum chamber temperature and acceleration. Note that between \SI{300}{\micro\Hz} and \SI{50}{\milli\Hz}, the~amplitude varies by a factor of nearly 5, suggesting that a linear regression would not be suitable for this data set. The~transfer function is not plotted below \SI{300}{\micro\Hz} because the low number of averages causes large uncertainties. The~shaded area represents the estimated error bars of the~spectrum.}
\label{fig:TempTransferFunction}
\end{figure}

The amplitude of the transfer function between the vacuum chamber temperature and the resonator's acceleration after removing laser frequency and barometric pressure noise is shown in Figure~\ref{fig:TempTransferFunction}. We observe a frequency-dependent transfer function amplitude ranging from \SI{2e-3}{\meter\second^{-2}\kelvin^{-1}} to \SI{1e-2}{\meter\second^{-2}\kelvin^{-1}}. Similar to laser frequency noise, the~transfer function presented in Figure~\ref{fig:TempTransferFunction} can be used to estimate the temperature stability required for the resonator to be limited by thermal motion. At~\SI{10}{\milli\Hz}, the amplitude of this transfer function is on the order of \SI{10}{\milli\meter\per\second\squared\per\kelvin}, while the thermal acceleration noise at this frequency is approximately \SI{4e-10}{\meter\per\second\squared\per\sqrt{\Hz}}. This suggests that a thermally limited acceleration measurement would require an environmental temperature stability of \SI{4e-8}{\kelvin\per\sqrt{\Hz}}. This requirement being very improbable to achieve, we instead investigated the precise mechanism through which temperature coupled into our setup and reduced it to relax the temperature stability we needed to~reach. 

The typical transfer function one might expect for temperature coupling is one that resembles a low-pass filter~\cite{nofrarias2013subtraction}, which was not fully observed in our setup. This behavior could be due to the fact that temperature was a ubiquitous effect, present everywhere in the setup. The~complicated transfer function could be composed of several thermal effects occurring simultaneously in different parts of the experiment. The~exact mechanism causing temperature fluctuations to couple into our setup in this manner is currently under~investigation.

\begin{figure}[htbp] 
\centering
\includegraphics[width=\linewidth]{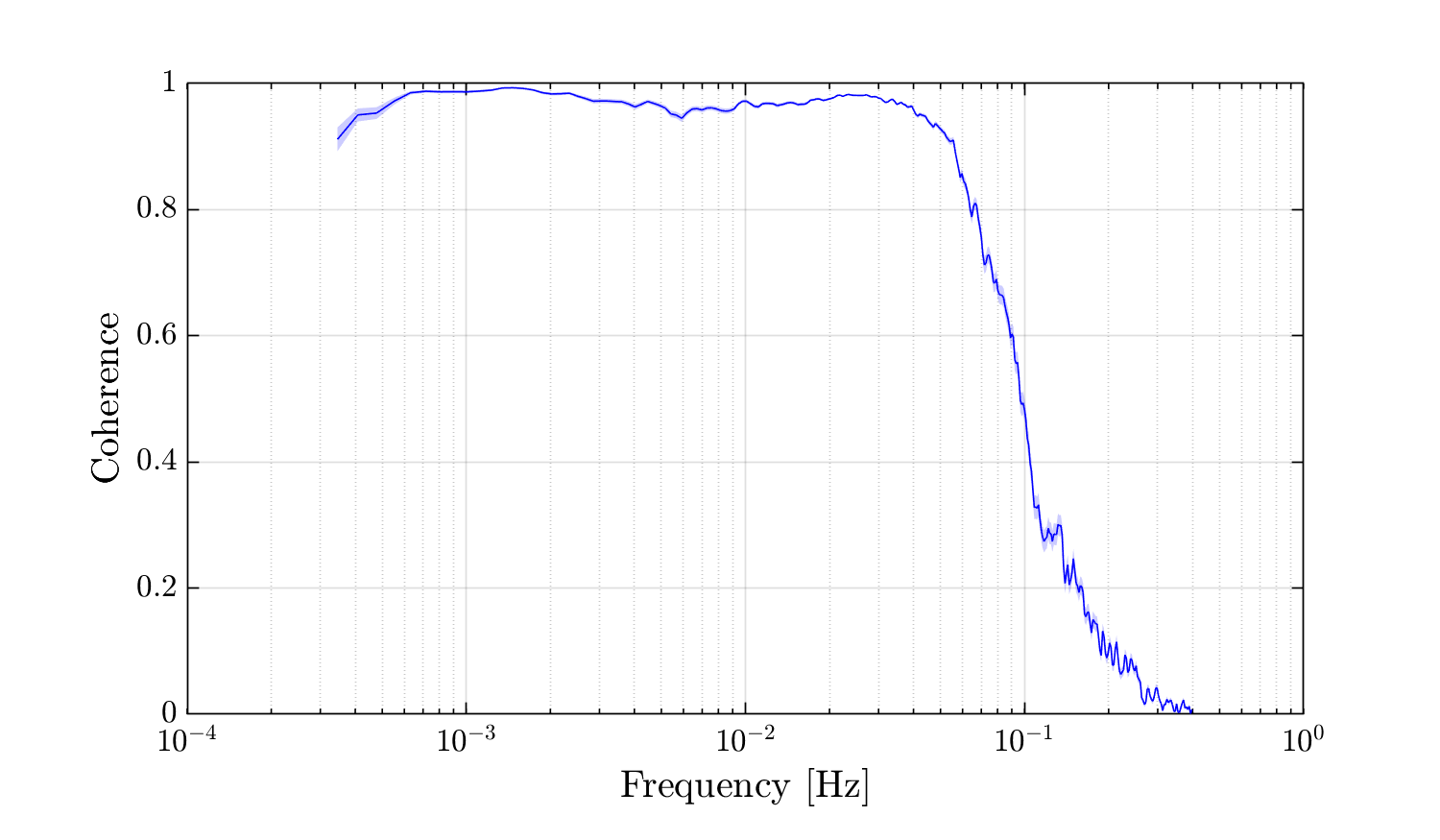}
\caption{The magnitude of the coherence between our vacuum chamber temperature and our resonator's acceleration data. The~coherence is greater than 0.8 between \SI{0.3}{\milli\Hz} and \SI{60}{\milli\Hz}, indicating that temperature is a significant noise source. The~coherence function is not plotted below \SI{300}{\micro\Hz} because the low number of averages causes large uncertainties. The~shaded area represents the estimated error bars of the~spectrum.}
\label{fig:TempCoherence}
\end{figure}

We also calculated the uncertainty in the transfer function, which depends on the transfer function amplitude, the~coherence between vacuum temperature and acceleration $C_{Ta}(\omega)$, and~the number of averages in each frequency bin~\cite{bendat1978statistical}. In~bins where either the coherence between the two time series was poor or the number of averages was low, the~variance in the transfer function was large. The~coherence, shown in Figure~\ref{fig:TempCoherence}, was estimated by taking the ratio of the magnitude of the cross-spectral density, $|S_{Ta}(\omega)|^{2}$, and~the product of the power spectral densities $S_{aa}(\omega)$ and $S_{TT}(\omega)$ \cite{bendat1978statistical}:
\begin{equation}
    C_{Ta}(\omega) = \frac{|S_{Ta}(\omega)|^{2}}{S_{aa}(\omega)S_{TT}(\omega)} \label{eq:C}
\end{equation}

The coherence between the vacuum chamber temperature and the acceleration data was larger than 0.8 between \SI{300}{\micro\Hz} and \SI{60}{\milli\Hz}, indicating that temperature fluctuations were a dominant noise source in this frequency band. Above~\SI{60}{\milli\Hz} there was a sharp decrease in the coherence, explaining the corresponding decrease in transfer function amplitude and the increase in the transfer function~uncertainty.

The temperature-induced acceleration noise was calculated by taking the FFT of the temperature data, applying the transfer function, and~then taking the inverse FFT to return to the time domain. The~inverse transform step of this process, while not strictly necessary if one wants frequency-domain results~\cite{bendat1978statistical}, is useful for visualizing the data and identifying the next limiting noise~source.

\begin{figure*}[htbp] 
\centering
\includegraphics[width=\linewidth,trim={0cm 1.25cm 0cm 1.25cm},clip]{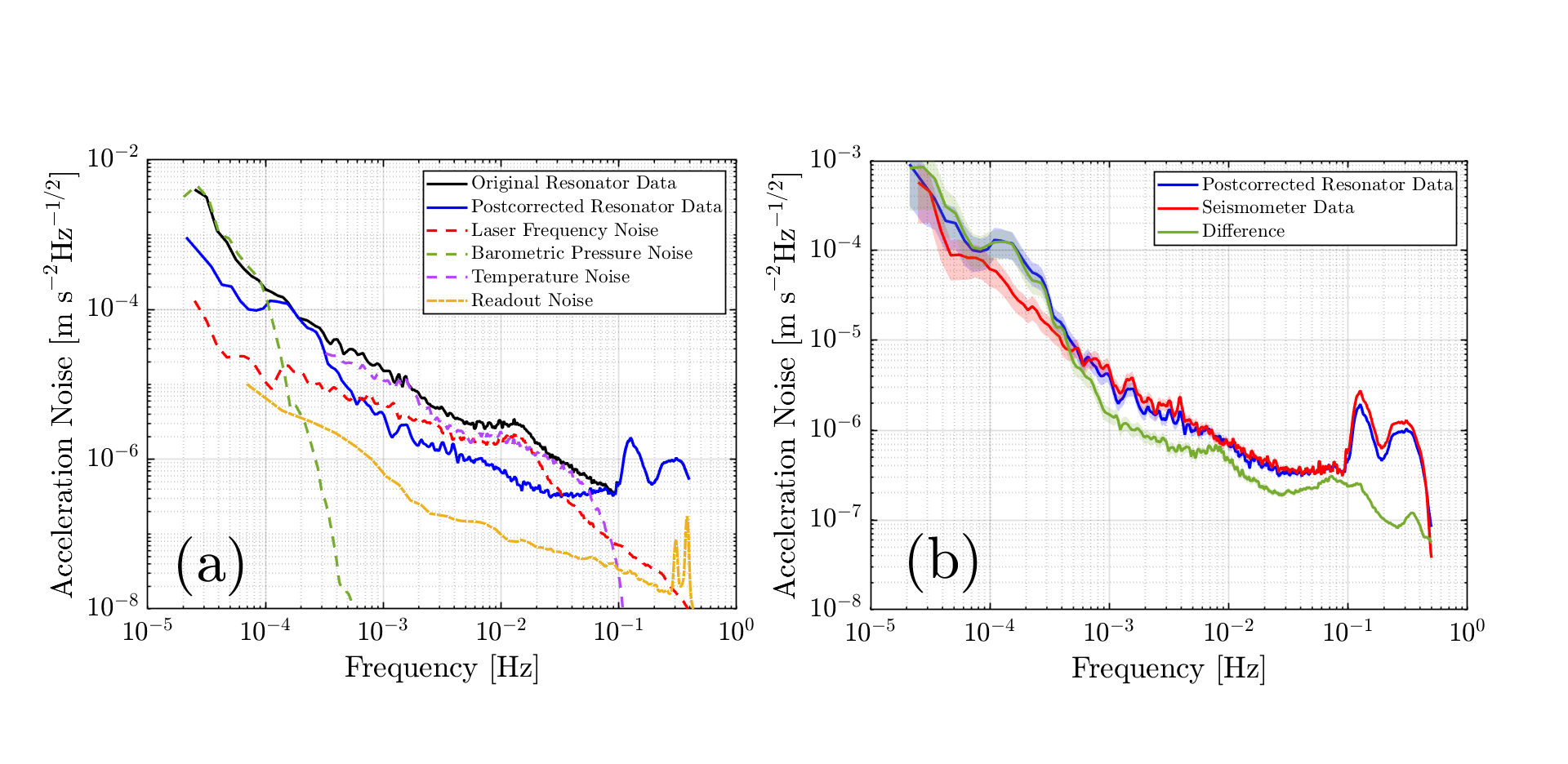}
\caption{(\textbf{a}) The amplitude spectral density of a seismic measurement taken with our optomechanical accelerometer alongside the noise contributions from laser frequency fluctuations, barometric pressure, and~vacuum chamber temperature. Shown also is the residual after removing all three noise sources, demonstrating a significant reduction in noise from \SI{0.2}{\milli\Hz} to \SI{100}{\milli\Hz} and below \SI{100}{\micro\Hz}. (\textbf{b}) A comparison of the postcorrected resonator data to the seismometer data. The~good agreement between the two devices now extends down to \SI{1}{\milli\Hz}. The~shaded areas represent the estimated error bars of the~spectra.}
\label{fig:NoiseBudget}
\end{figure*}

The noise breakdown for this measurement is shown in Figure~\ref{fig:NoiseBudget}a as well as the residual after removing the laser frequency, pressure, and~temperature noise. This postcorrection is compared to the measured interferometric readout noise, from~which we see that the resonator's acceleration data are above the acceleration noise of the interferometer. This suggests that the noise we are observing is a real signal with seismic or environmental origins. Figure~\ref{fig:NoiseBudget}b also shows the difference between our optomechanical accelerometer data and the seismometer data. We observe that this difference is larger than the readout-induced acceleration noise, suggesting that the residual is limited by noise not originating from the optical readout. Our data do not show any more significant coherences between our acceleration and environmental monitoring. As~such, we consider the possibility that the residual noise in our accelerometer data was, at~least partially, introduced through postcorrection. For~example, if~the self-noise of our temperature sensors was on the order of \SI{100}{\micro\kelvin\per\sqrt{\Hz}} at \SI{10}{\milli\Hz}, that self-noise would be added to our acceleration measurements during postcorrection at a level of approximately \SI{2e-7}{\meter\second^{-2}\per\sqrt{\Hz}}, ultimately restricting our final sensitivity. This highlights the importance of investigating and characterizing these noise sources, as reducing their physical coupling into the experiment offers better results than decohering our data in~postcorrections. 

\begin{figure}[htpb] 
\centering
\includegraphics[width=\linewidth]{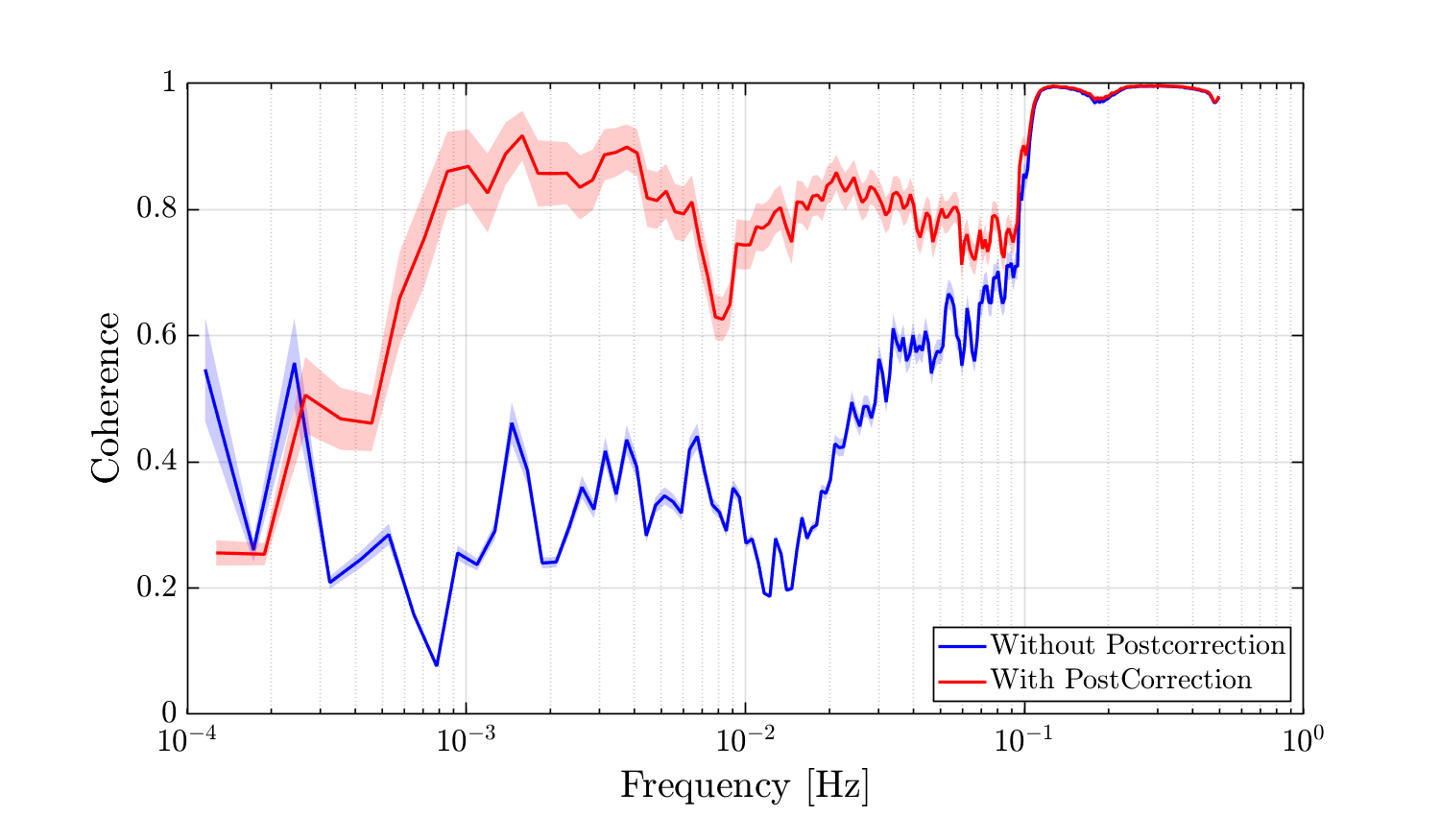}
\caption{The magnitude of the coherence between our postcorrected resonator data and seismometer data both before and after removing environmental noises. It can be seen that by removing temperature, pressure, and~laser frequency fluctuations from the resonator data, the~coherence above \SI{1}{\milli\Hz} improves significantly. The~shaded areas represent the estimated error bars of the~spectra.}
\label{fig:coherence}
\end{figure}

\begin{figure*}[htpb] 
\centering
\includegraphics[width=\linewidth]{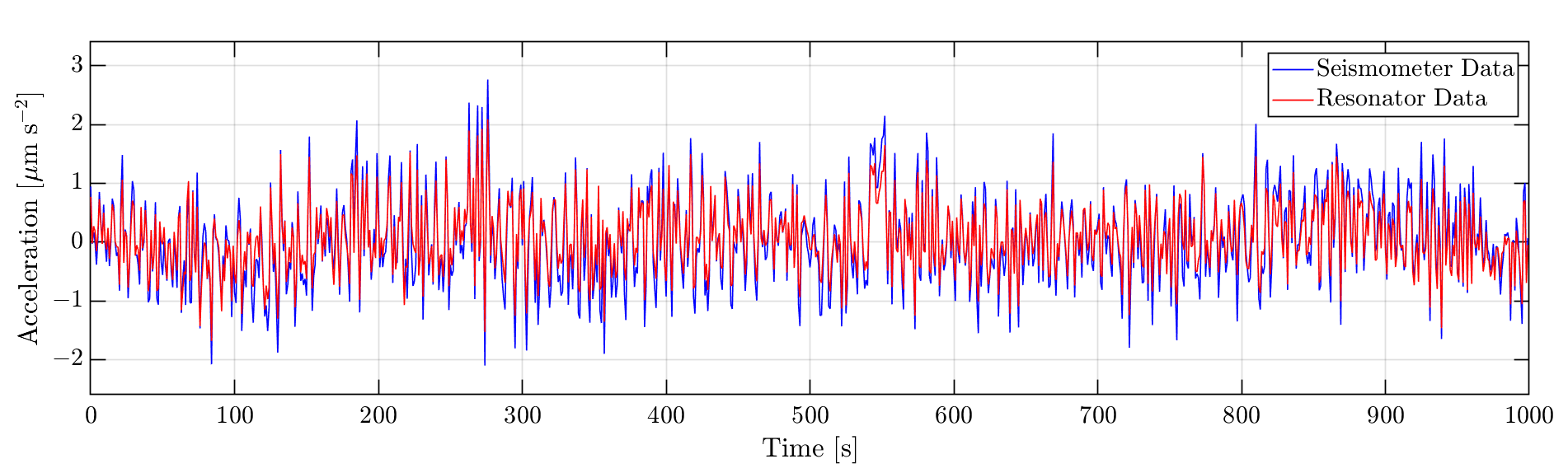}
\caption{A 1000-s-long subset of the acceleration time series obtained by our optomechanical accelerometer and seismometer, in~which we see excellent agreement between the two time series. Both traces are high-pass filtered above \SI{0.8}{\milli\Hz} for better comparison of the frequency band with high~coherence.}
\label{fig:TimeDomainComparison}
\end{figure*}

Furthermore, we find that after removing the temperature and barometric pressure fluctuations, the~good agreement between our resonator and seismometer data extends down to \SI{1}{\milli\Hz}, which can be visualized in frequency-space in Figure~\ref{fig:NoiseBudget}a and in the coherence between the two data sets in Figure~\ref{fig:coherence}. The~bump in the postcorrected resonator data between \SI{100}{\micro\Hz} and \SI{400}{\micro\Hz} is likely residual barometric pressure noise that could not be removed due to the low Nyquist frequency of the pressure measurement. Moreover, we can visually compare the data from both accelerometers in the time domain to demonstrate the agreement between the sensors. In~Figure~\ref{fig:TimeDomainComparison}, we plotted a 1000 s segment of both data streams and in doing so, we observed excellent agreement in both the magnitude and phase of the two devices. These comparisons validated our resonator's ability to detect seismic noise above \SI{1}{\milli\Hz}.

\section{Discussion}
\subsection{Comparative Technology~Assessment}
\begin{figure}[htbp] 
\centering
\includegraphics[width=1.1\linewidth]{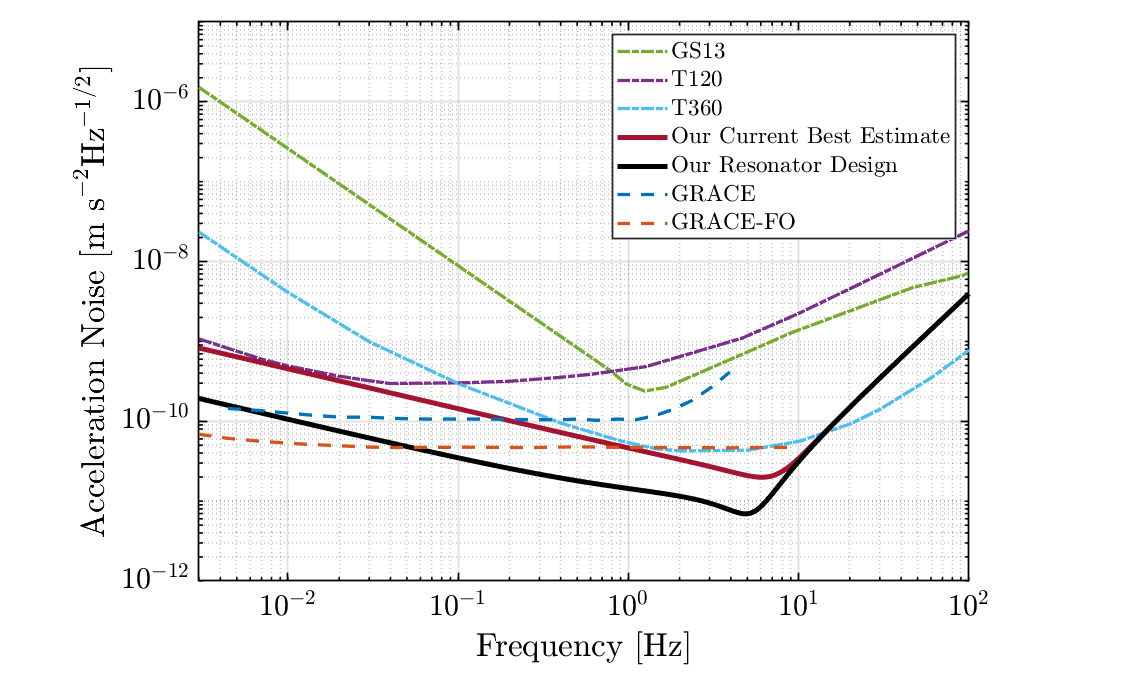}
\caption{The projected noise floor of our optomechanical accelerometer compared to those of other technologies, including the GRACE\,\cite{flury2008precise} and GRACE-FO~\cite{BANDIKOVA2019623} accelerometers and the Geotech S-13, Trillium 120, and~Trillium 360 seismometers~\cite{PrivateCommunication}. Two traces are plotted for the noise floor of our accelerometer: our current best estimate using our experimental value of the $Q$-factor, and~our resonator design using a value of $Q$ calculated using models for the loss mechanisms of fused~silica.}
\label{fig:TechComparison}
\end{figure}

In Section~\ref{sec:NoiseFloor}, we calculated the noise floor of our optomechanical accelerometer using estimates of the optical readout noise and the thermal motion associated with our experimentally measured $Q$-factor. This noise floor can be compared to existing technologies to assess the viability of our accelerometer for different applications. In~Figure~\ref{fig:TechComparison}, we plotted our estimated noise floor against the GRACE~\cite{flury2008precise} and GRACE-FO~\cite{BANDIKOVA2019623} accelerometers, as~well as several ground-based seismometers including the Geotech S-13, the~Nanometrics Trillium 360, and~the Nanometrics Trillium 120, which is the one we utilized in our laboratory for the comparison measurements presented in Section~\ref{sec:Measurement}~\cite{PrivateCommunication}. We observe that our accelerometer is anticipated to have a competitive, if~not lower, acceleration noise than the ground-seismometers, suggesting that our device will be useful for seismometry and ground-based geodesy studies. Moreover, our accelerometer has a mass of \SI{58.2}{\gram}, and~we anticipate the realization of a highly compact and lightweight system compared to existing systems. Because~of this, our optomechanical accelerometer is smaller, more portable, and~better-suited for field work than these commercial~technologies.

When compared to the GRACE and GRACE-FO accelerometers, our optomechanical accelerometer is competitive at higher-frequencies around \SI{1}{\Hz} but~is expected to be noisier at lower-frequencies. However, current mass change measurements taken by satellite gravity recovery missions are limited primarily by temporal aliasing errors, not accelerometer noise~\cite{wiese2011expected}. Hence, despite this slightly higher noise floor at low frequencies, we expect our optomechanical accelerometers to be able to provide meaningful data for science observations. Rather, with~the advantages exhibited by our optomechanical accelerometer, including a lower weight, more compact form, and~the ability to be tested on-ground, our accelerometer is still valuable for satellite geodesy~missions.

\subsection{Planned~Developments}
\label{sec:future}
\begin{figure*}[htpbp]
\centering
\includegraphics[width=\linewidth]{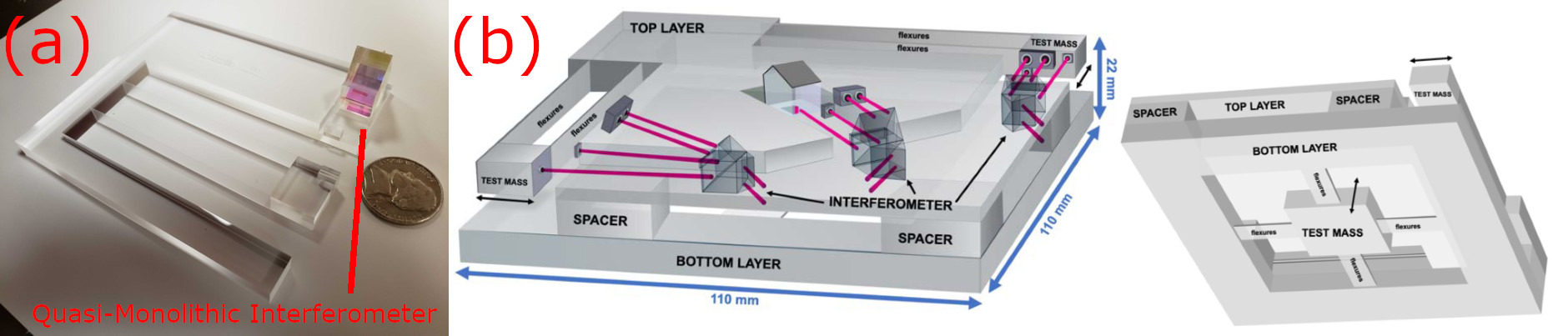}
\caption{(\textbf{a}) A visualization of our quasi-monolithic interferometer integrated onto the wafer of our resonator. A~US nickel is included for scale. (\textbf{b}) A rendering of a triaxial accelerometer concept. This design consists of two layers: the top containing resonators oscillating in the $x$ and $y$-axes and the bottom with a resonator operating along the $z$-axis.}
\label{fig:PlannedResDev}
\end{figure*}

In the measurements presented in this article, we encountered several noise sources, such as instabilities in the laser frequency and the optical readout, limiting our ability to detect low-frequency signals. Future measurements will be conducted with a frequency-stabilized laser and a compact quasi-monolithic interferometer to improve our sensor's low-frequency sensitivity. This interferometer operates on the same principles as our current readout. It consists of prisms and beamsplitters bonded together and small enough that they can be integrated onto the wafer of our resonator, miniaturizing our current experimental setup while offering better mechanical stability. A~photo of this interferometer is shown in Figure~\ref{fig:PlannedResDev}a.

Other improvements that can be made to reduce the measurement noise include thermally isolating the system to dampen temperature fluctuations as well as investigating noise due to tilt-to-length coupling and nonlinear optical path length~differences.

Furthermore, we will experimentally measure the thermal noise floor of our accelerometer by performing a Huddle test. The~test consists of placing two identical resonators close to each other. The~seismic noise would couple into both devices coherently. After~removing that correlated seismic noise, the~remaining noise will be the uncorrelated noise originated from the resonators~themselves.

These developments and a better understanding of low-frequency environmental noise sources will allow for other tests to characterize our acceleration sensing capabilities, e.g.,~to monitor the lunar and solar tidal acceleration~changes.

Finally, there are several extensions of the work presented in this manuscript that would make this optomechanical accelerometer technology ready for use onboard a space geodesy satellite. Specifically, we are currently developing appropriate mounts and launch-lock mechanisms for the accelerometer that facilitate its deployment and protect the dynamic test mass during installation and launch. This cage should include a method of securing the test mass and flexures to avoid damage while not in use, such as during launch when significantly higher accelerations and shock are expected in contrast to science operation. To~this end, the~higher-order violin modes of the flexures in our accelerometer should be investigated in greater depth to ensure compatibility with the vibrational loads present during~launch.

Moreover, a~triaxial optomechanical accelerometer could be constructed from a series of resonators operating along different axes, which is required for observing all noninertial perturbations while in orbit. A~concept design of such a sensor is depicted in Figure~\ref{fig:PlannedResDev}b, which has a volume of $\SI{110}{\milli\meter}\times\SI{110}{\milli\meter}\times\SI{22}{\milli\meter}$ and a mass of \SI{0.282}{\kilo\gram}, not including the masses of the optical readout, which should be low overall as each quasi-monolithic interferometer assembly weighs only \SI{4.5}{\gram}. This design is a quasi-monolithic assembly of two layers of fused silica resonators that are separated by spacers. The~top layer holds two orthogonal resonators identical to the one presented in Section~\ref{sec:ResonatorDesign} operating along the $x$ and $y$ axes. Because~these resonators share the same topology as the one presented in this paper, they will have the same noise floors shown in Figure~\ref{fig:ThermalNoise}. Furthermore, on the top layer is a cutout that houses the optical readouts for all three axes. An~interferometer reaches the test mass of the bottom layer through a hole in the top layer covered by a~pentaprism.

The bottom layer contains a resonator that measures out-of-plane accelerations along the $z$-axis. However, due to limitations in fabrication, a different topology is required for measuring the out-of-plane motion. To~this end, we have developed a preliminary design for the bottom layer resonator, which consists of a \SI{12}{\gram} rectangular prism test mass supported by flexures on four sides. These flexures have cross sections of
 $\SI{8}{\milli\meter}\times\SI{0.1}{\milli\meter}$ and are oriented such that the test mass oscillates along the $z$-axis. Tuning the lengths of the flexures as well as the test mass allows for a natural frequency less than \SI{10}{\Hz}, which improves the resonator's low-frequency sensing capabilities. Like the resonator presented in Section~\ref{sec:ResonatorDesign}, we require all higher order modes to have resonances larger than the natural mode by an order of magnitude. In~the topology we have chosen for the bottom layer resonator, however, tip-and-tilt modes can have a low enough frequency to allow cross talk between the modes of the~resonator.

\begin{figure}[htbp]
\centering
\includegraphics[width=\linewidth]{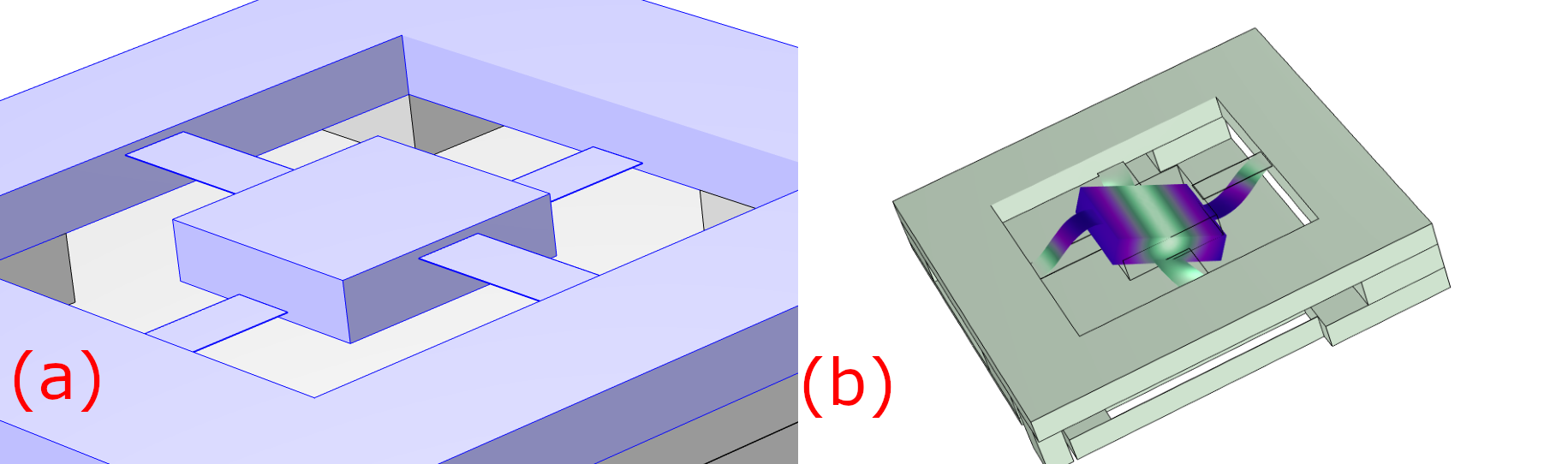}
\caption{(\textbf{a}) A close-up of the $z$-axis test mass and its flexures. Note that the flexures are not at the midpoint of the wafer but~have different offsets to increase the stiffnesses and frequencies of tip-tilt modes. (\textbf{b}) The tip-tilt mode of the $z$-axis resonator as seen from the underside of the three-axis resonator. The~colors indicate the total displacement, with~light green being the stationary and purple being the largest~displacement.}
\label{fig:ZAxisResonator}
\end{figure}

Figure~\ref{fig:ZAxisResonator}a visualizes this tip-and-tilt mode. To~combat this, we offset the flexure heights relative to the bottom of the wafer. Two flexures are \SI{1}{\milli\meter} above the bottom of the wafer and the other two are \SI{1}{\milli\meter} below the top of the wafer. Having the test mass and flexures connected at multiple points along the $z$-axis increases the frequency of this tip-tilt mode to above the 10$\times$ threshold. Figure~\ref{fig:ZAxisResonator}b depicts a close-up of the test mass and flexures to show this~offset.

With the aim of developing a triaxial accelerometer with equal noise floors along all three axes, an~optimization of the bottom layer resonator must be performed to ensure it has comparable noise to the top layer resonators. This optimization includes adjusting the test mass and cross section to lower its resonance and reduce its thermal motion while keeping the stress in the flexures within a safe operating~range. 

\section{Conclusions}
In this paper, we demonstrated that a compact and lightweight optomechanical accelerometer etched from a monolithic wafer of fused silica with dimensions \linebreak of 
 \mbox{$\SI{90}{\milli\meter}\times\SI{80}{\milli\meter}\times\SI{6.6}{\milli\meter}$} could detect seismic noise above \SI{1}{\milli\Hz} in good agreement with commercial seismometer technologies. This device had a test mass that was suspended mechanically by thin flexures, allowing it to be tested on the ground where some electrostatic devices cannot. As~such, we proposed the use of this technology onboard future space geodesy missions, as~well as in ground-based planetary and geodesy applications, for~measuring noninertial~disturbances. 

Although a direct measurement of the accelerometer's noise floor was not possible due to our testing environment being flooded with signals, there is presently no evidence of noise sources that would prevent us from observing the self-noise of our accelerometer with a sufficiently quiet test bed. The~anticipated thermal acceleration noise was approximately \SI{5e-11}{\meter\second^{-2}\per\sqrt{\Hz}} at \SI{1}{\Hz}, making this technology competitive with the accelerometers that have flown on space geodesy missions such as GRACE and GRACE-FO. As~such, our optomechanical accelerometer is expected to be suitable for satellite geodesy missions, among~other~applications.

The resonator's displacement was measured optically with a heterodyne interferometer and a \SI{1064}{\nano\meter} laser that was not frequency-stabilized. Below~\SI{100}{\milli\Hz}, the acceleration data obtained by our resonator were dominated by laser frequency noise as well as temperature and pressure fluctuations. These noise sources can be partially removed by careful environmental monitoring, significantly increasing the coherence between our resonator and a commercial seismometer. Future works will incorporate improvements to our optical readout, laser source, and~environmental monitoring to further enhance our acceleration~measurements. 

\subsection*{Author contributions}
Conceptualization, Adam Hines and Felipe Guzm\'an; Formal analysis, Adam Hines and Jose Sanjuan; Funding acquisition, Felipe Guzm\'an; Investigation, Adam Hines, Andrea Nelson, Yanqi Zhang and Guillermo Valdes; Methodology, Adam Hines, Andrea Nelson, Yanqi Zhang, Guillermo Valdes, Jeremiah Stoddart and Felipe Guzm\'an; Software, Jeremiah Stoddart; Supervision, Guillermo Valdes and Felipe Guzm\'an; Validation, Adam Hines; Visualization, Adam Hines; Writing – original draft, Adam Hines; Writing – review \& editing, Andrea Nelson, Yanqi Zhang, Guillermo Valdes, Jose Sanjuan and Felipe Guzm\'an. All authors have read and agreed to the published version of the manuscript.

\subsection*{Funding}
This research was funded by: National Geospatial-Intelligence Agency (NGA) (grant HMA04762010016),
National Science Foundation (NSF) (grant PHY-2045579),
National Aeronautics and Space Administration (NASA) (grants 80NSSC20K1723 and 80NSSC22K0281), and Jet Propulsion Laboratory (JPL) (contract 1677619).

\subsection*{Conflict of Interest}
The authors declare no conflict of interest. The funders had no role in the design of the study; in the collection, analyses, or interpretation of data; in the writing of the manuscript; or in the decision to publish the results.

\bibliography{References}

\end{document}